\def\papertitle{Neural modeling of magnetic tape recorders}
\def\paperauthorA{Otto Mikkonen}
\def\paperauthorB{Alec Wright}
\def\paperauthorC{Eloi Moliner}
\def\paperauthorD{Vesa Välimäki}

\documentclass[twoside,a4paper]{article}
\usepackage{etoolbox}
\usepackage{dafx_23}
\usepackage{amsmath,amssymb,amsfonts,amsthm}
\usepackage{euscript}
\usepackage[T1]{fontenc}
\usepackage[utf8]{inputenc}
\usepackage{ifpdf}
\usepackage[english]{babel}
\usepackage{caption}
\usepackage{subcaption} 
\usepackage{color}
\usepackage{lipsum}     
\usepackage{siunitx}    
\DeclareSIUnit \dBFS {\dB \text{FS}} 
\usepackage{booktabs}   
\usepackage{array}      
\newcolumntype{C}[1]{>{\centering\arraybackslash}m{#1}} 
\usepackage[export]{adjustbox} 

\input glyphtounicode
\ninept

\newcounter{numauth}
\newcounter{listcnt}
\newcommand\authcnt[1]{\ifdefined#1 \stepcounter{numauth} \fi}

\newcommand\addauth[1]{
\ifdefined#1 
\stepcounter{listcnt}
\ifnum \value{listcnt}<\value{numauth}
\appto\authorslist{, #1}
\else
\appto\authorslist{~and~#1}
\fi
\fi}
\authcnt{\paperauthorB}
\authcnt{\paperauthorC}
\authcnt{\paperauthorD}
\authcnt{\paperauthorE}
\authcnt{\paperauthorF}
\authcnt{\paperauthorG}
\authcnt{\paperauthorH}
\authcnt{\paperauthorI}
\authcnt{\paperauthorJ}
\def\authorslist{\paperauthorA}
\addauth{\paperauthorB}
\addauth{\paperauthorC}
\addauth{\paperauthorD}
\addauth{\paperauthorE}
\addauth{\paperauthorF}
\addauth{\paperauthorG}
\addauth{\paperauthorH}
\addauth{\paperauthorI}
\addauth{\paperauthorJ}

\usepackage{times}

\newif\ifpdf
\ifx\pdfoutput\relax
\else
   \ifcase\pdfoutput
      \pdffalse
   \else
      \pdftrue
\fi

\usepackage[pdftex,
 pdftitle={\papertitle},
 pdfauthor={\authorslist},
 pdfsubject={Proceedings of the 26th International Conference on Digital Audio Effects (DAFx23)},
 colorlinks=false, 
 bookmarksnumbered, 
 pdfstartview=XYZ 
]{hyperref}
\usepackage[hypcap=true]{caption}
\title{\papertitle}

\affiliation{
\paperauthorA, \,
\paperauthorB, \,
\paperauthorC  \, and 
\paperauthorD \sthanks{\scriptsize This work was supported by the Nordic Sound and Music Computing Network (NordForsk project number 86892). }
}
{
\href{https://www.aalto.fi/en/aalto-acoustics-lab}{Acoustics Lab}, Department of Information and Communications Engineering \\
Aalto University, Espoo, Finland \\
{\tt {firstname.lastname@aalto.fi} }
}

\begin{document}
\ifpdf 
 \DeclareGraphicsExtensions{.png,.jpg,.pdf}
\else  
 \DeclareGraphicsExtensions{.eps}
\fi


\maketitle

\begin{abstract}
 The sound of magnetic recording media, such as open-reel and cassette tape recorders, is still sought after by today's sound practitioners due to the imperfections embedded in the physics of the magnetic recording process.
 This paper proposes a method for digitally emulating this character using neural networks.
 The signal chain of the proposed system consists of three main components: the hysteretic nonlinearity and filtering jointly produced by the magnetic recording process as well as the record and playback amplifiers, the fluctuating delay originating from the tape transport, and the combined additive noise component from various electromagnetic origins.
 In our approach, the hysteretic nonlinear block is modeled using a recurrent neural network, while the delay trajectories and the noise component are generated using separate diffusion models, which employ U-net deep convolutional neural networks.
 According to the conducted objective evaluation, the proposed architecture faithfully captures the character of the magnetic tape recorder.
 The results of this study can be used to construct virtual replicas of vintage sound recording devices with applications in music production and audio antiquing tasks.
\end{abstract}

\section{INTRODUCTION}\label{sec:intro}

\begin{figure}[t]
 \centerline{
  \includegraphics[width=0.8\columnwidth]{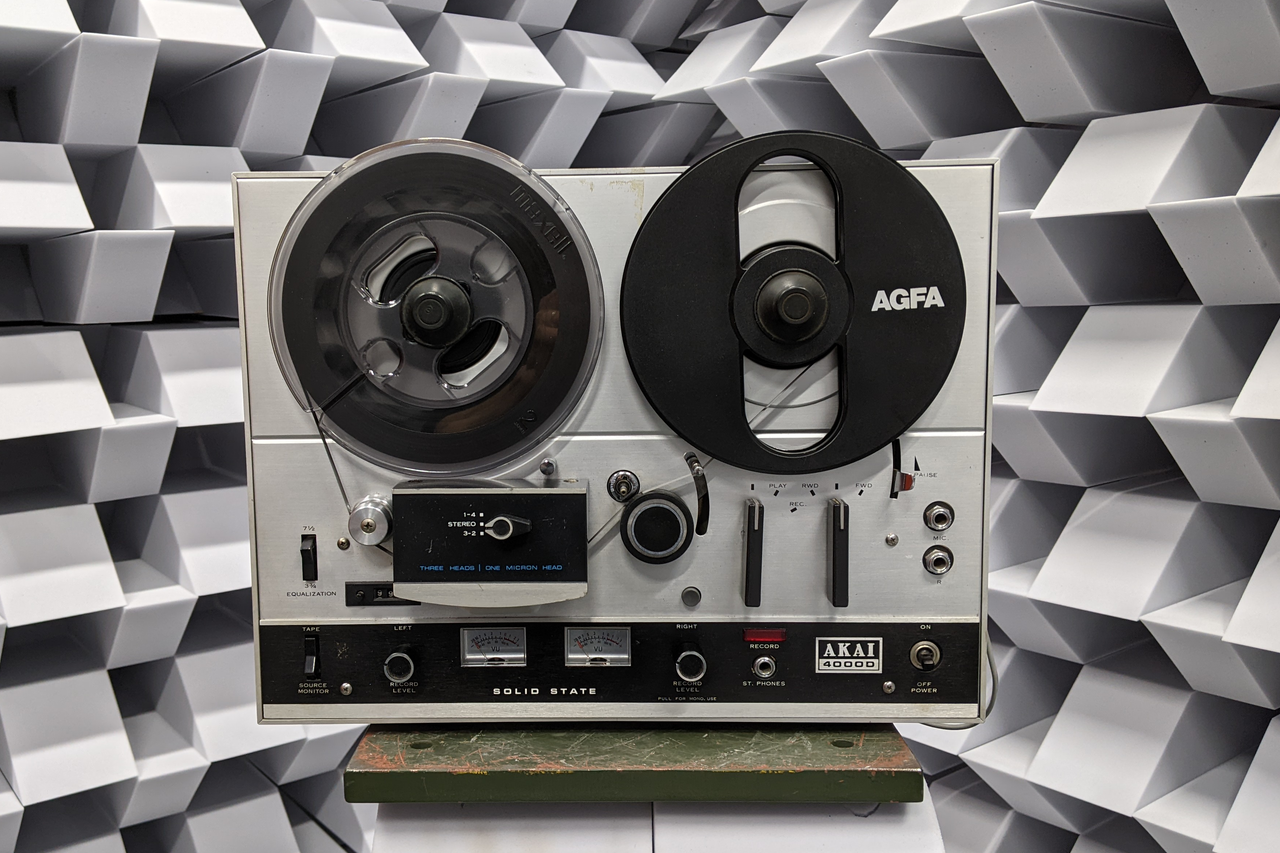}
 }
 \vspace{-1mm}
 \caption{\label{fig:intro-akai-4000d}{\it Akai 4000D reel-to-reel tape recorder.}}
 \vspace{-5mm}
\end{figure}

Magnetic recording has had a profound impact on the history of recorded music, providing a dramatic leap in the quality of the stored audio in comparison to the earlier direct-to-disk techniques. The advances in magnetic recording grounded practices such as multitrack and sound-on-sound recording within the industry, and as the technology matured and became cheaper, allowed for entire generations of professional and amateur musicians alike to experiment with these powerful production techniques. 
Reel-to-reel tape recorders, an example of which is shown in Fig.~\ref{fig:intro-akai-4000d}, have been largely replaced by digital recording techniques for sound capture and reproduction.
However, the idiosyncrasies of the magnetic recording process are now used as a creative effect. This paper studies the imperfections of the magnetic recording process and emulates them digitally and with neural networks.  

Virtual analog (VA) modeling is an area of digital signal processing with a rich lineage in the past decades, aiming at modeling analog audio devices and making the emulations available as software \cite{zolzer_virtual_2011}. The techniques used for the modeling are traditionally divided into white-box, grey-box, and black-box methods, depending on the type of information used as the basis for the task. White-box techniques use exact knowledge of the underlying circuits to reconstruct the physics in the digital domain in order to match the observed behavior. Black-box techniques use observations collected from the target as the basis and optimize a general-purpose method to replicate the behavior. In grey-box modeling, a combination of white- and black-box methods is used. Over the last decade, advances in deep learning have given rise to their increased application also for VA modeling, with the techniques capable of exhibiting state-of-the-art performance in a number of different tasks \cite{wright_real-time_2019, moliner_realistic_2022, massi_deep_2023}.

While the physics underlying magnetic recording has been studied thoroughly in the past \cite{bertram_theory_1994, camras_magnetic_1998}, it has been applied for the purpose of digitally emulating the recording process only recently \cite{chowdhury_real-time_2019}.
Pertinent to this topic is the lineage of work related to the modeling of tape delay devices, which has been covered more extensively \cite{arnardottir_digital_2008, zolzer_virtual_2011, kaloinen_neural_2022}.
In our work, we build up from earlier literature on emulating the sound of the magnetic tape, and propose a grey-box system for the emulation task. We consider the sound of the tape recorder to be build up of the nonlinear hysteretic magnetization of the tape medium, the filtering produced by the recording and playback heads, a fluctuating delay induced by the imperfections in the tape transport mechanism, the tape hiss, as well as the subtle nonlinearities and filtering of the input/output amplifiers. The proposed system uses a hybrid array of modern deep-learning techniques as the backbone for modeling these different aspects of the character.

\begin{figure*}[ht]
 \centering
 \begin{subfigure}[b]{0.685\linewidth} 
  \centering
  \adjincludegraphics[trim={{0.00\width} 0 {0.00\width} 0},clip,width=1.0\columnwidth]{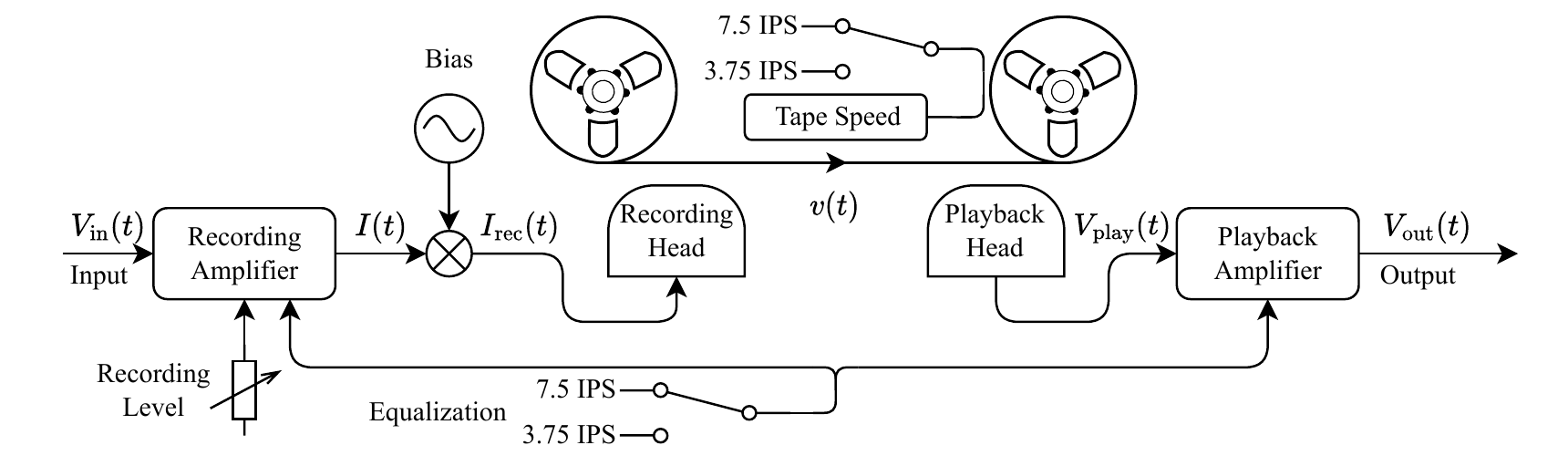}
  \caption{\label{fig:theory-reel-to-reel}{}}
 \end{subfigure}%
 \begin{subfigure}[b]{0.315\linewidth} 
  \centering
  \adjincludegraphics[trim={{0.00\width} 0 {0.00\width} 0},clip,width=1.0\columnwidth]{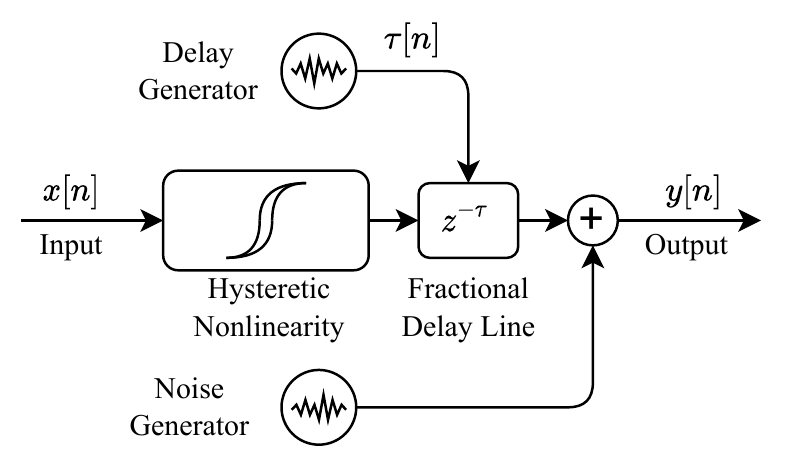}
  \caption{\label{fig:modeling-system}{}}
 \end{subfigure}
 \vspace{-5mm}
 \caption{\it (a) Real and (b) VA system block diagrams.}
 \vspace{-5mm}
\end{figure*}

The rest of this paper is organized as follows. 
Sec.~\ref{sec:theory} reviews the background theory regarding the effects of magnetic recording. 
Sec.~\ref{sec:modeling} gives an overview of the system proposed for the modeling.
Sec.~\ref{sec:data} provides details concerning the data used to evaluate the proposed method, while Sec.~\ref{sec:training} describes the training process for the different network architectures.
The experimental procedure is divided into two sections depending on the type of data used for the evaluation: in Sec.~\ref{sec:experiment-1}, toy data collected from a virtual tape machine is used, while Sec.~\ref{sec:experiment-2} uses data from a real machine. 
Sec.~\ref{sec:conclusions} concludes the work.

\section{MAGNETIC RECORDING}\label{sec:theory}
The block diagram of a typical magnetic recorder is shown in Fig.~\ref{fig:theory-reel-to-reel}.
The system consists of a recording amplifier, a recording head, the moving tape medium, a playback head, and a playback amplifier.
The following paragraphs briefly discuss each of these components and their contribution to the overall character, based on earlier literature \cite{chowdhury_real-time_2019, bertram_theory_1994, camras_magnetic_1998, zolzer_virtual_2011, arnardottir_digital_2008, kaloinen_neural_2022}.

\begin{figure}[t]
 \centerline{
  \adjincludegraphics[trim={0 {0.05\height} 0 0},clip,width=1.0\columnwidth]{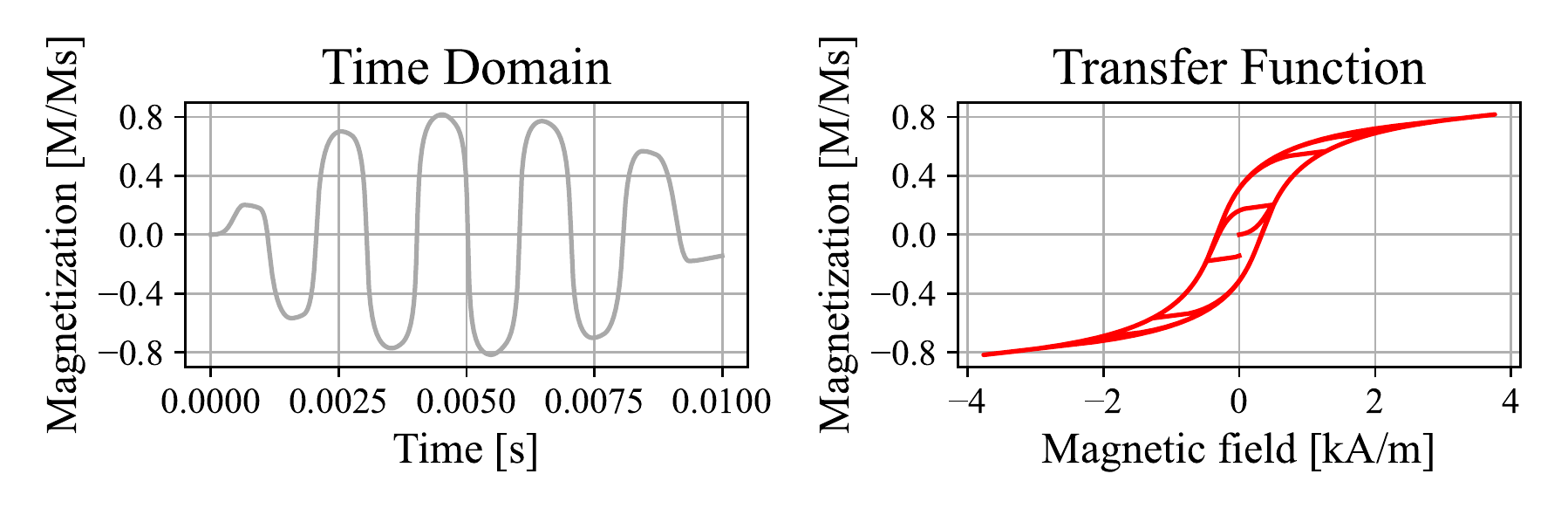}
 }
 \vspace{-2mm}
 \caption{\label{fig:theory-hysteresis}{\it Tape magnetization nonlinearity exhibits hysteresis.}}
 \vspace{-3mm}
\end{figure}

The recording head takes the input current from the recording amplifier, and produces a spatial magnetic field determined by both the properties of the recording head and the magnitude of the input current. When the moving magnetic tape is exposed to this field, the magnetic dipoles in the substrate take the form of the field, which is retained as it passes the volume of the field. Since the characteristic magnetization of the tape takes a hysteretic nonlinear shape, shown in Fig.~\ref{fig:theory-hysteresis}, a high-frequency bias is added to the input signal to reduce the nonlinearity produced by the hysteresis.

\begin{figure}[t]
 \centerline{
  \adjincludegraphics[trim={0 {0.525\height} 0 0},clip,width=1.0\columnwidth]{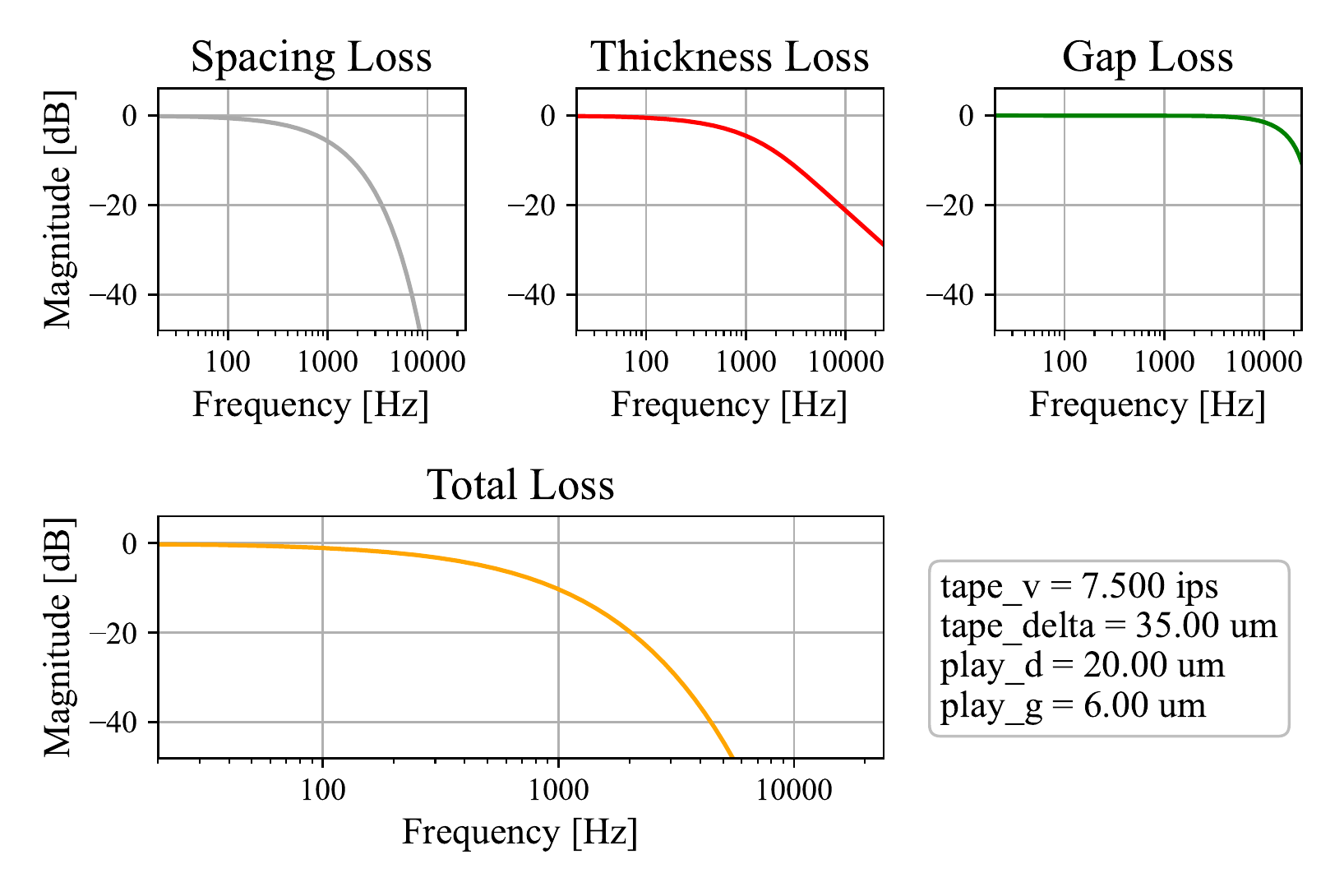}
 }
 \vspace{-2mm}
 \caption{\label{fig:theory-playback}{\it Three components of playback losses.}}
 \vspace{-3mm}
\end{figure}

As the tape moves past the playback head, the changing magnetic field induces a current within its internal coil, restoring part of the stored signal into electrical form. This recovery of the signal is a spatial integration over the magnetized volume of tape, which leads to tape-speed-dependent filtering, shown in Fig.~\ref{fig:theory-playback}. The filtering consists of components related to the spacing between the head and the tape, the tape thickness, and the playback head gap.

To counterweight the spatial filtering induced by the playback head, the recording and playback amplifiers are used as pre- and post-filtering stages. This also serves to maximize the dynamic range of the tape and condition the input and output signals of the system. In our reference design shown in Fig.~\ref{fig:intro-akai-4000d}, these amplifiers are implemented with cascaded transistor circuits, producing additional low-order harmonic distortion in the processed signal.

\begin{figure*}[ht]
 \centering
 \begin{subfigure}[t]{0.5\linewidth}
  \captionsetup{margin={11mm, 0cm}} 
  \centering
  \adjincludegraphics[trim={0 {0.666666\height} 0 0},clip,width=1.0\linewidth]{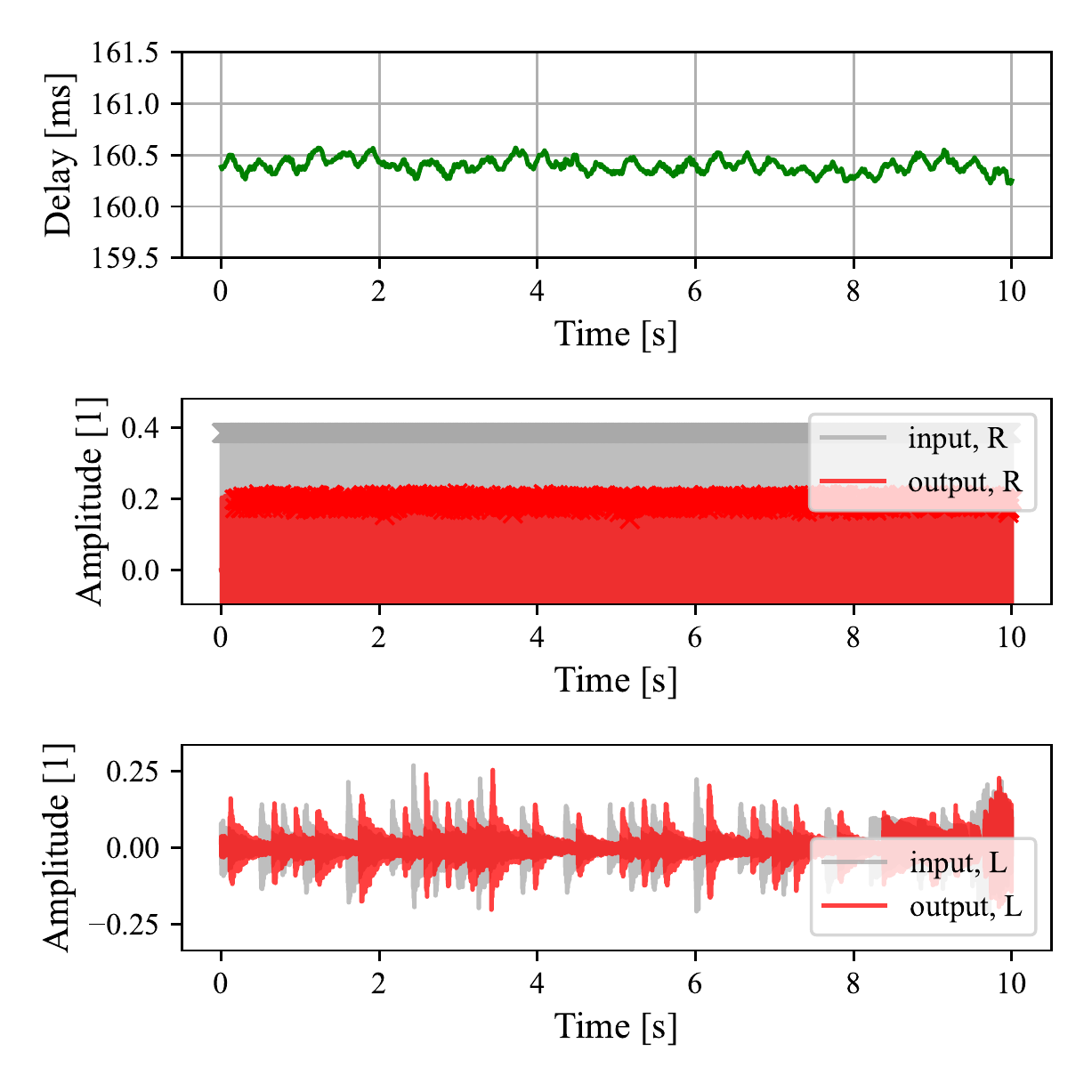}
  \vspace{-5mm}
  \caption{\label{fig:theory-trajectories}{}}
 \end{subfigure}%
 \begin{subfigure}[t]{0.5\linewidth}
  \centering
  \captionsetup{margin={11mm, 0cm}} 
  \adjincludegraphics[trim={0 {0.35\height} 0 {0.333333\height}},clip,width=1.0\linewidth]{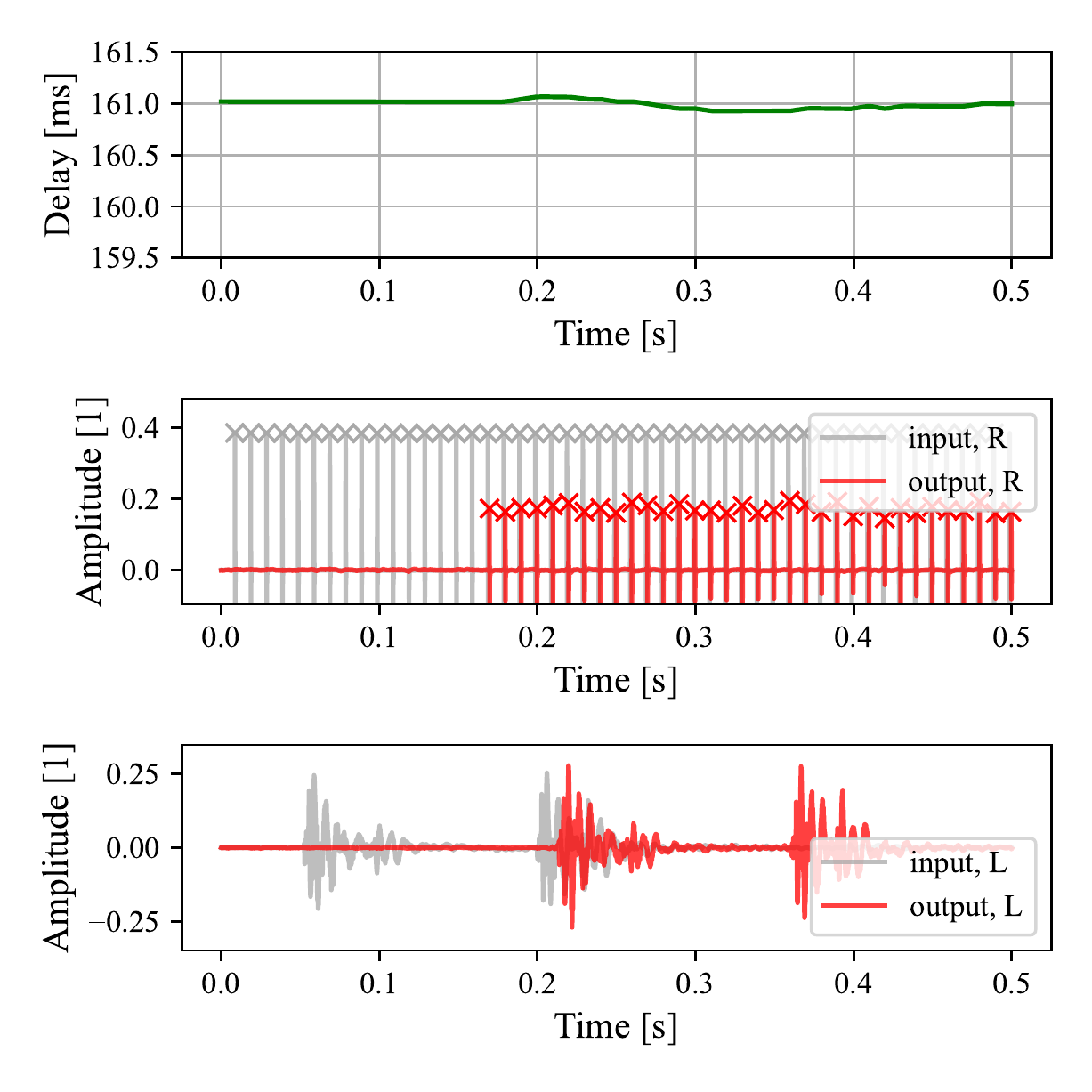}
  \vspace{-5mm}
  \caption{\label{fig:pulse-detection}{}}
 \end{subfigure}
 \vspace{-3mm}
 \caption{ \label{fig:deltraj}{\it (a) Delay trajectories and (b) measurement signal.}}
 \vspace{-3mm}
\end{figure*}

The tape movement speed is not perfectly constant, due to small fluctuations produced by imperfections in the tape transport mechanics.
These imperfections include cyclical components produced by the moving parts in the transport mechanism as well as stochastic behavior, shown in Fig.~\ref{fig:theory-trajectories}, in the form of a delay trajectory. These inconsistencies in the movement can be heard as small fluctuations in pitch, known as \textit{wow} and \textit{flutter}.

Magnetic tape recorders generate noise artifacts throughout the recording and playback stages, which contribute to the distinctive qualities of the resulting audio signal. The noises could be originated by multiple sources, such as the playback equipment, the magnetic particles in the tape coating, modulation noises during recording, or surface asperities, among others \cite{daniel_tape_1972}.

\section{METHODS}\label{sec:modeling}
We employ a grey-box model, shown in Fig.~\ref{fig:modeling-system}, inspired by the block diagram of the target device. The signal path for the proposed system consists of three components: 1) a hysteretic nonlinearity for modeling the magnetic recording process lumped with the record/playback amplifier responses, 2) a time-varying delay line controlled by a delay trajectory generator, and 3) an additive noise component. In the following, details concerning these components as well as the capture of delay trajectories are given. 

\subsection{Delay Trajectory Retrieval}\label{ssec:modeling-pulse-detection}
The fluctuating time delay between the record and playback heads is captured using a pulse train-based measuring technique \cite{arnardottir_digital_2008, zolzer_dafx_2011, chowdhury_real-time_2019}.
The measurement signal consists of a train of unit impulses spaced $T = 1/f$ apart, where $f$ is the repetition frequency. This signal is played through the target device, and the locations of the input and output pulses are compared to determine the time delay between the heads as it fluctuates over time.
The frequency $f$ of the pulse train determines the sampling rate for the captured trajectories
and in our experiments $f = \SI{100}{\hertz}$ was used \cite{arnardottir_digital_2008, kaloinen_neural_2022}.
An example of a measured delay trajectory, as well as the measurement signal at the input and output of a studied target are illustrated in Fig.~\ref{fig:deltraj}.


\subsection{Lumped Nonlinearities}\label{ssec:modeling-nonlinearity}
The lumped effects include the hysteretic nonlinearity of the magnetization process, the spatial filtering of the playback head, as well as the low-order distortion components and filtering originating from the record and playback amplifiers.
We choose a recursive neural network (RNN) architecture for modeling these aspects, consisting of a gated recurrent unit and a linear output layer \cite{wright_real-time_2019}. The model details can be found in the original research article. The choice of the model stems from the stateful nature of the recurrent unit, which we hypothesize as being helpful in learning the hysteresis shape. Three training schemes for training the RNN are considered: two supervised and one adversarial, explained in the following subsections.

\begin{figure*}[ht]
 \centering
 \begin{subfigure}[t]{0.45\linewidth}
  \centering
  \captionsetup{margin={2mm, 0mm}} 
  \adjincludegraphics[trim={{0.05\width} 0 {0.2\width} 0},clip,width=1.0\columnwidth]{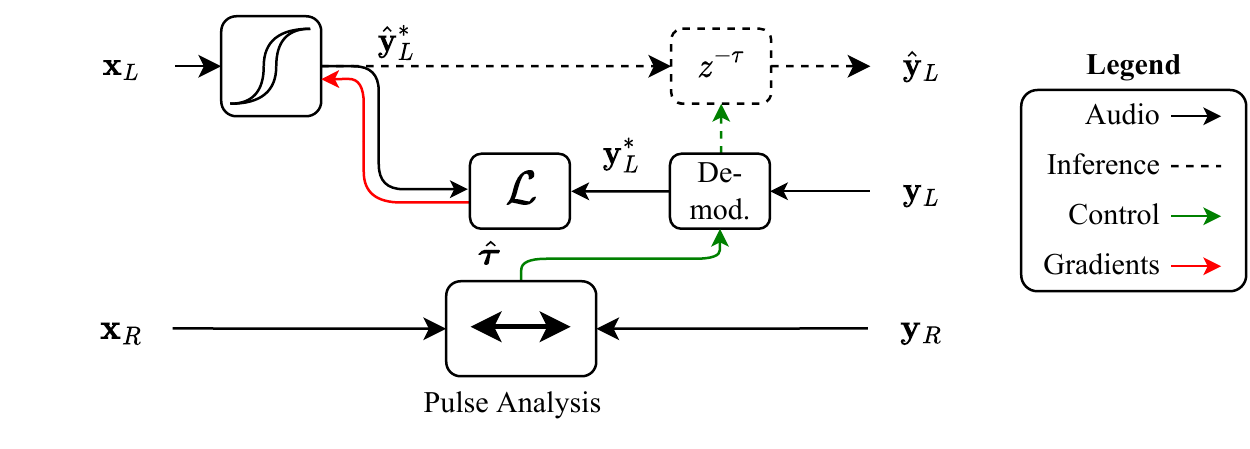}
  \vspace{-6mm}
  \caption{\label{fig:modeling-nonlinearity-approach-1}{}}
 \end{subfigure}%
 \begin{subfigure}[t]{0.1\linewidth}
  \centering
  \adjincludegraphics[trim={{0.81\width} {0.0\height} 0 {0.0\height}},clip,width=1.0\columnwidth]{images/nonlinearity-approach-1-no_data.pdf}
  \vspace{-6mm}
 \end{subfigure}%
 \begin{subfigure}[t]{0.45\linewidth}
  \centering
  \captionsetup{margin={4mm, 0mm}} 
  \adjincludegraphics[trim={{0.05\width} 0 0 0},clip,width=1.0\columnwidth]{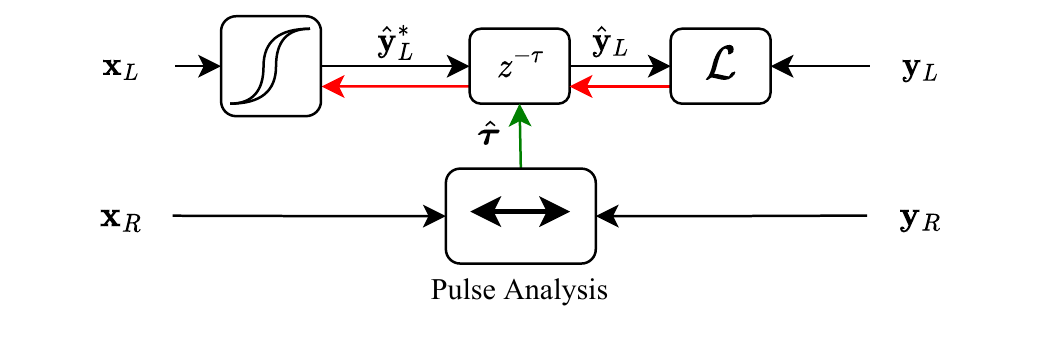}
  \vspace{-6mm}
  \caption{\label{fig:modeling-nonlinearity-approach-2}{}}
 \end{subfigure}
 \vspace{-2mm}
 \caption{\label{fig:modeling-nonlinearity-approaches}{\it Supervised approaches for training the nonlinearity: (a) Demodulation and (b) differentiable delay line.}}
\end{figure*}

\subsubsection{Supervised Approaches}\label{ssec:modeling-signal_path-supervised-1}
For the supervised approaches, we exploit the stereo nature of most tape recorders. We construct stereo training signals with the audio content on the left channel embedded with the measurement pulse train signal on the right. Using the pulse detection algorithm to construct the delay trajectories for each individual audio segment, the time evolution of the tape medium can be captured and used to restore the alignment of the targets and model predictions.

We implement and compare two approaches for restoring the alignment, shown in Fig.~\ref{fig:modeling-nonlinearity-approaches}.
In the first approach (Fig.~\ref{fig:modeling-nonlinearity-approach-1}), inspired by Kaloinen's work \cite{kaloinen_neural_2022}, we use the captured delay trajectory $\boldsymbol{\tau}$ to demodulate the target segment $\mathbf{y}^*_L = \text{demod}(\mathbf{y}_L, \boldsymbol{\tau})$, where $^*$ denotes a signal which has been demodulated or has not had a delay line applied to it. After this procedure, the demodulated signal and the raw output from the nonlinear block become aligned and we compute the loss $\mathcal{L}(\hat{\mathbf{y}}^*_L, \mathbf{y}^*_L)$. In this scenario, the gradients flow only through the nonlinear block, and the time-varying delay line is only added during inference.

In the second approach, shown in Fig.~\ref{fig:modeling-nonlinearity-approach-2},
we use the captured delay trajectory $\boldsymbol{\tau}$ to delay the raw output from the nonlinear block
$\hat{\mathbf{y}}_L[n] = \hat{\mathbf{y}}^*_L[n - \boldsymbol{\tau}[n]]$
in order to compute the loss directly as $\mathcal{L}(\hat{\mathbf{y}}_L, \mathbf{y}_L)$. In this approach, the gradients flow also through the delay line, which needs to be differentiable.
Linear interpolation was used to implement the continuously variable delay line, and the implementation was originally included in the Magenta DDSP codebase \cite{engel_ddsp_2020-1}. Care must be taken to avoid non-differentiable rounding operations such as the floor operator.

The delay line is initially filled from left to right with integers representing delay-line length, starting from $N$ and decreasing with a step size of 1, with the final value being 0. The desired delay-line length is subtracted from each element of this vector and the abs operator is applied. Each element of this vector is subtracted from 1, and finally, the rectified linear unit (ReLU) function is applied, resulting in a vector where all indices are filled with zeros, except for the indices with indexes immediately above and below the desired delay-line length. This vector can then be multiplied element-wise with a buffer of previous input values, and the sum of this operation produces the output of the delay line.

\subsubsection{Adversarial Approach}\label{ssec:modeling-nonlinearity-training-2}
An alternative approach using an adversarial training method is also proposed. This allows for the modeling of monophonic tape recorders. In this method, measured or synthesized delay lines are applied to the RNN model output. Instead of training the model using a supervised loss function, which requires the delay line to be known, an adversarial loss can be used such that the delay line applied does not have to match the actual delay line applied in the training data. In this case, a discriminator model is trained to distinguish between real examples of processed audio from the target dataset, and synthetic examples which are produced using our modeling approach. The discriminator model and training procedure used are identical to those used earlier by Wright et al.~\cite{wright_adversarial_2023}. The discriminator receives a time-frequency representation of the audio as input, and consists of a stack of 1D convolutional layers, with the first layer treating the frequency bins as an input channel. The discriminator and the tape model are trained adversarially using the hinge loss function.


\subsection{Noise Generator}\label{ssec:modeling-noise}

The background noise component is modeled using a diffusion probabilistic model \cite{ho_denoising_2020}, in a similar fashion to previous work from Moliner and Välimäki~\cite{moliner_realistic_2022}. A diffusion model is used as a data-driven universal approximator of the probability distribution of all the additive disturbances that are introduced during the recording, magnetization, and playback processes. 
The model can be trained with recorded silent passages containing only the background textures produced by the reel-to-reel machine. 
By reversing a diffusion process, white Gaussian noise segments are progressively morphed into noises from the training data distribution.
Based on the assumption that the noises are additive, the generated noise samples are added to the output signal as a final step. 

Although we have adopted the main concept from a previous work \cite{moliner_realistic_2022} as a basis for our research, there are significant deviations in the technical details of our approach due to our use of more recent developments on diffusion models. We adopt some of the design choices from Karras et al.~\cite{karras_elucidating_2022}, including the ordinary differential equation (ODE) formulation, the neural network preconditioning, the training objective, and the noise schedule parameterization, the latter being a Variance Exploding noise schedule \cite{song_score-based_2021}.

\subsection{Trajectory Generator}\label{ssec:modeling-trajectories}

Given the stochastic nature of the delay trajectories underlined in Sec.~\ref{sec:theory}, in this work, they are modeled using a probabilistic generative model. 
Similar to Sec.~\ref{ssec:modeling-noise}, a diffusion model is also used to for this task. In this case, the model is trained to emulate the distribution of the measured delay trajectories.

\section{DATA COLLECTION}\label{sec:data}
This section provides details concerning the data used in the experimental procedure, including the compiled datasets.
The compiled datasets are made available in the accompanying webpage \footnote[1]{\url{http://research.spa.aalto.fi/publications/papers/dafx23-neural-tape/}}.

As input data, we use a fraction of the inputs from SignalTrain \cite{hawley_profiling_2019}
for training the nonlinear block. The dataset consists of short musical passages representing varying genres played using various instruments, together with synthetic measurement signals, sampled at $\SI{44.1}{\kilo\hertz}$. A total of $60 \text{ min}, 20 \text{ min, and } 15 \text{ min}$ of audio is used for training, validation, and testing, respectively.

\subsection{Toy Data}\label{sec:toydata}
We test the modeling architecture using synthetic data, generated via wrapping the VST instance of CHOWTape \cite{chowdhury_real-time_2019}, a white-box modeled tape machine, using Pedalboard \footnote[2]{\url{https://github.com/spotify/pedalboard}}.
During the generation, the VST instance is set to $16\times$ oversampling using an $8$-iteration Newton-Rhapson solver for the tape hysteresis ODE, which are the highest quality settings available.
To make sure the virtual tape is sufficiently saturated, the tape drive, tape saturation, and tape bias are set to $(0.75, 0.75, 0.0) \in [0,1]$, respectively. The timing parameters---the flutter depth, wow depth, and wow variance---are set to $(0.75, 0.75, 1.0) \in [0,1]$, respectively. We turn off any additional processing from the VST
which did not appear in the original research article.
Two datasets are collected for the experiments with the toy data: one with only the tape effects enabled and the timing effects disabled, and one with both of the effects enabled.

\subsection{Real Data}
The real data for evaluating the modeling architecture was collected using an Akai 4000D open-reel tape recorder (Fig.~\ref{fig:intro-akai-4000d}). The device is a $\tfrac{1}{4}$ inch, four-track, three-head, stereo recorder from the 1970s, capable of running at $3 \tfrac{3}{4}$ and $7 \tfrac{1}{2}$ 
inches per second (IPS) and using discrete transistor circuitry for the input/output amplifiers. The data was collected using two types of magnetic tape: a Maxell low-noise/high-output tape and a Scotch low-noise tape from the 1970s. 
An RME Fireface UCX audio interface was used for recording and playback.

Using a three-head recorder allows for simultaneous recording and playback to and from the tape, allowing the fluctuating time delay between the record and playback heads to be captured. In practice, we connect a stereo line feed from the audio interface into the line inputs of the tape recorder, set monitoring to \textit{TAPE}, and record both the stereo line feed from the tape recorder, as well as a loopback signal from the interface outputs back to its inputs. The recording level of the device was set such that when monitoring the signal entering the tape (monitoring set to \textit{INPUT}), a \SI{0}{\dB FS} signal from the interface is just below the clipping threshold of the record and playback amplifiers.

The collected data is divided into datasets used for training the lumped nonlinearities, the delay trajectory generator, and the noise generator. We collected two versions of each dataset using the (tape branch, tape speed) configurations (MAXELL,  $7 \tfrac{1}{2}$ IPS) and (SCOTCH,  $3 \tfrac{3}{4}$ IPS). While initially the same datasets were intended to be used for training both the nonlinear block and the delay trajectory generator, studying the extracted trajectories from the lumped nonlinearity datasets revealed that the accuracy of the trajectory generator would be severely limited by the considered sampling rate of \SI{44.1}{\kilo\hertz}, and thus we collected separate high-resolution datasets at \SI{192}{\kilo\hertz} for the trajectory generator, consisting of the same upsampled audio. 
Finally, to train the noise generator, two datasets consisting of only the hiss captured from the line feed of the tape recorder were collected using the original sampling rate of \SI{44.1}{\kilo\hertz}.

\section{IMPLEMENTATION DETAILS}\label{sec:training}
This section provides details concerning implementing the different components in the modeling architecture, including modeling training and the loss functions used. 

\subsection{Supervised Approaches}
The first two approaches use supervised training to optimize the weights of the nonlinear block, aligning the target and output segments using the proposed methods shown in Fig.~\ref{fig:modeling-nonlinearity-approaches}. The RNN is trained using truncated back-propagation through time (TBPTT) \cite{elman_finding_1990}, allowing the RNN state to initialize before tracking the gradients. For the first approach, the RNN state is initialized for $1024$ steps, and for the second, the initialization length is determined by taking the next power of two of the maximum delay length in samples encountered in the training dataset.
We use a hidden size of $64$ as preliminary experiments indicated that increasing the hidden size beyond this did not bring an improvement in model performance.
We use Adam with the default hyperparameters as implemented in PyTorch as the optimizer. We use a learning rate of $1\times10^{-3}$ and reduce it with a factor of $0.75$ every time the validation loss has not improved for $10$ epochs. A batch size of $32$ is used for all the experiments and the models are trained using a graphical processing unit for \SI{4}{\text{hours}}.

To compute the prediction discrepancy against the target output for the supervised approaches, the error-to-signal ratio (ESR) loss is used \cite{damskagg_deep_2019}. Details concerning the loss function can be found in the paper by Damskägg et al.~\cite{damskagg_deep_2019}.


\subsection{Adversarial Approach}
For the adversarial training method, TBPTT was also used. As the time-varying delay line is initially filled with zeros, the initialization stage is run for a number of steps equal to the maximum measured delay-line length. This ensures that the delay line is filled with real values during TBPTT. A segment length of 2 seconds was used during training, with parameter updates being carried out every 16384 samples. The longer TBPTT length was used as the discriminator model uses a time-frequency representation of the signal as input. As such, longer input lengths increase the frequency resolution that is seen by the discriminator model.

The discriminator and RNN model are alternately trained using the hinge loss function described by Kumar et al.~\cite{kumar_melgan_2019}. A batch size of 16 was used during training. During validation, paired data was used to evaluate the RNN model, with the measured delay line being applied at the output of the RNN. Training was run for 50 epochs, with a multi-resolution log spectral magnitude loss being used to select the best performing model weights. 

\subsection{Noise Generator}\label{sec:training_noise}

We train our diffusion models following the recommendations by Karras et al.~\cite{karras_elucidating_2022}.
Considering that the standard deviation of the recorded data is, approximately, $\sigma_\text{data}=8\times10^{-4}$, the noise schedule is defined so that, during sampling, the Gaussian noise level decreases logarithmically across the reverse diffusion process from $\sigma_\text{max}=0.1$ (completely masking the data) to $\sigma_\text{min}=5\times10^{-5}$ (perceptually insignificant). During training, the model is trained with the L2 preconditioned objective from \cite{karras_elucidating_2022}, where the noise level is sampled randomly with a LogUniform distribution. The model is trained with the Adam optimizer with the default momentum hyperparameters and a learning rate of $2 \times 10^{-4}$. An exponential moving average of the weights with a decay factor of 0.999 is tracked during training and used as the final inference model.

A standard time-domain convolutional U-Net is used as the backbone deep neural network architecture, which is conditioned on a noise level embedding that allows weights to be shared across different noise levels. The total number of parameters adds to 127k, which is relatively low when compared to standard practice for diffusion models, but has proven qualitatively to be enough for this particular use-case. Inference is performed with a denoising diffusion implicit model (DDIM) \cite{song_denoising_2021} sampler with a subtle amount of stochasticity, which allows for a trade-off between sampling speed and quality by adjusting the number of sampling steps.  We use 16 steps in our experiments, but we observe that this number can be reduced down to 6 without a significant quality loss.

The noise generator model is trained with samples of 1.5-s at the sampling rate of 44.1 kHz. However,  given that the architecture is fully-convolutional, the segment size could be freely adapted during inference. In addition, arbitrarily long sequences can be generated by applying a chunked autoregressive sampling strategy, where separate frames are concatenated and inter-frame coherence can be ensured by applying a zero-shot outpainting technique \cite{ho_video_2022}.

\subsection{Trajectory Generator}
The delay trajectory generator is trained in a very similar way as the noise generator, specified in Sec.~\ref{sec:training_noise}, while only some implementation details concerning the characteristics of the data differ. The model is trained with 5.2-s segments at a sampling frequency of 100 Hz, defined by the measuring pulse frequency (see Sec.~\ref{ssec:modeling-pulse-detection}), resulting on segments of 512 samples. During training, every delay trajectory segment is mean-normalized, leaving only the local fluctuations from the average delay as the distribution to be modeled. The backbone neural network architecture is also a standard convolutional U-Net with 77k trainable parameters; we refer to the source code for further details.  The noise schedule design is motivated analogous to Sec.~\ref{sec:training_noise} and depends on the statistics of the dataset. 
For the toy data experiment (see Sec.~\ref{sec:experiment-1}), the approximated data standard deviation is $\sigma_\text{data}=6.8 \times 10^{-3}$, and the noise schedule is designed between $\sigma_\text{max}=0.5$ and $\sigma_\text{min}=1 \times 10^{-5}$. 
For the real data experiment (see Sec.~\ref{sec:experiment-2}), the data standard deviation is approximately  $\sigma_\text{data}=1 \times 10^{-4}$, and we found that $\sigma_\text{max}=0.01$ and $\sigma_\text{min}=1 \times 10^{-5}$ were a suitable design choice. 
The sampling is performed with a 10-step DDIM sampler \cite{song_denoising_2021}, but we observed that the number of discretization steps could be reduced down to only 4 with minimal qualitative differences.

\section{EXPERIMENT 1: TOY DATA}\label{sec:experiment-1}
This section presents experiments using synthetic data in two subsections. 
Sec.~\ref{ssec:experiment-1.1} studies the capability of the modeling method for the hysteretic magnetization of the tape without the fluctuating timing effects. 
Later, the timing effects are also enabled, allowing the evaluation of the three training schemes for the nonlinearity in Sec.~\ref{ssec:experiment-1.2} and the trajectory generator in Sec.~\ref{ssec:experiment-1.3}. The toy data does not contain a noise component, however. Audio examples for these experiments are available on the accompanying web page \footnotemark[1].

\subsection{Lumped Nonlinearities Only}\label{ssec:experiment-1.1}
To evaluate model performance, we use the ramped sine technique for the hysteresis \cite{holters_circuit_2016}. Additionally, we encourage readers to listen to the example predictions on the accompanying web page\footnotemark[1]. The ramped sine technique is especially useful here, since it evaluates both the deadzone effect resulting from an under-biased tape and the saturation at higher amplitudes.

\begin{figure}[t]
 \centerline{
  \adjincludegraphics[trim={0 {0.52\height} 0 10},clip,width=1.0\linewidth]{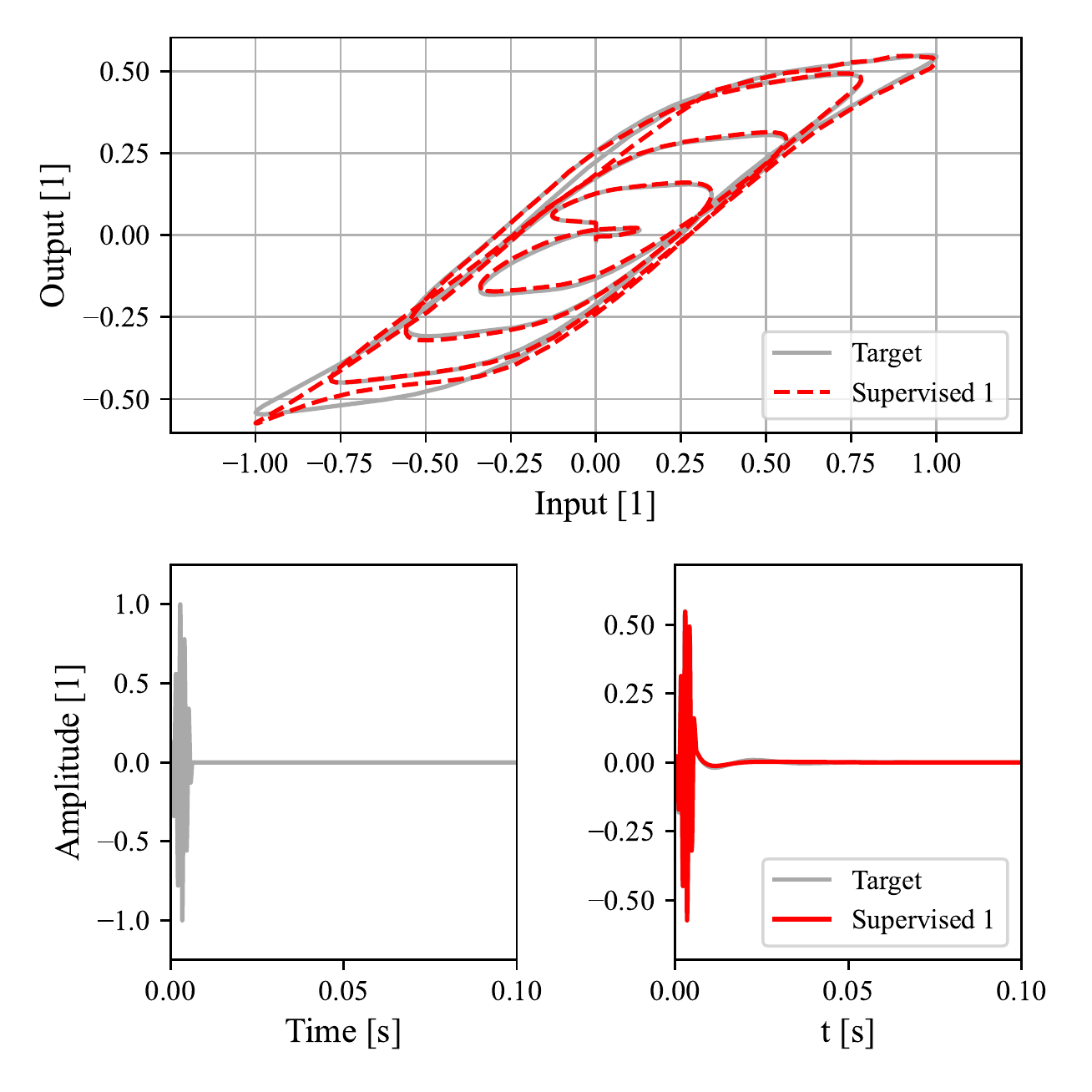}
 }
 \vspace{-2mm}
 \caption{\label{fig:exp1a-hysteresis}{\it Model hysteresis using toy data - Lumped nonlinearities only.}}
 \vspace{-3mm}
\end{figure}

The model hysteresis versus the target is shown in Fig.~\ref{fig:exp1a-hysteresis}. While the match is not perfect, the model clearly learns the shape of the hysteresis loop, including the deadzone effect, the saturation of the tape, as well as the loop width. Listening and comparing the model predictions against the target further validates this finding, 
confirming the suitability of the RNN architecture for modeling the type of nonlinearity.

\begin{figure}[t]
 \centerline{
  \adjincludegraphics[trim={0 {0.52\height} 0 10},clip,width=1.0\linewidth]{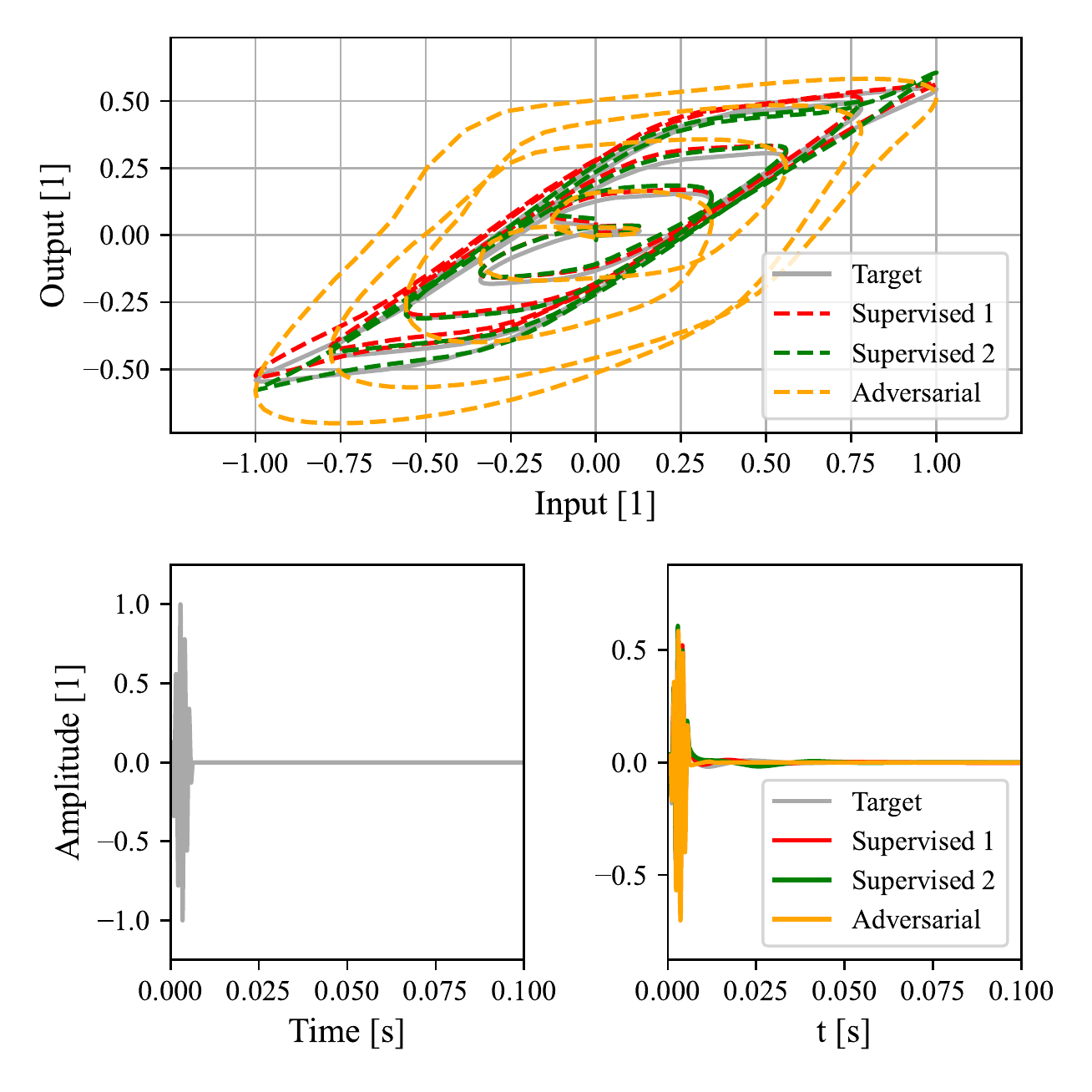}
 }
 \vspace{-2mm}
 \caption{\label{fig:exp1b-hysteresis}{\it Model hysteresis using toy data - Nonlinearities and timing effects.}}
 \vspace{-3mm}
\end{figure}
\begin{table}[b!]
 \vspace{-5mm}
 \caption{\itshape Toy Data: Nonlinearities and timing effects.}
 \vspace{-3mm}
 \centering
 \begin{tabular}{C{0.21\columnwidth} C{0.13\columnwidth} C{0.13\columnwidth} | C{0.13\columnwidth} C{0.13\columnwidth} }
  
  \toprule
                & \multicolumn{2}{c|}{$\mathcal{L}_\text{ESR}$} & \multicolumn{2}{c}{$\mathcal{L}_\text{STFT}$}                              \\
  Approach      & Demod.                                        & Delayed                                       & Demod.  & Delayed          \\\midrule
  Supervised I  & $0.031$                                       & $\mathbf{0.029}$                              & $0.645$ & $0.536$          \\
  Supervised II & --                                          & $\mathbf{0.029}$                              & --    & $\mathbf{0.488}$ \\
  Adversarial   & --                                          & $1.567$                                       & --    & $1.772$          \\
  \bottomrule
 \end{tabular}
 \label{tab:exp2}
\end{table}

\subsection{Lumped Nonlinearities and Timing Effects} \label{ssec:experiment-1.2}
The various training schemes are evaluated as in Sec.~\ref{ssec:experiment-1.1}, on top of which the model losses over the test set are compared. Since the adversarial models are not trained with a time-domain loss, we include a multi-resolution short-time Fourier transform (STFT) loss \cite{yamamoto_parallel_2020} in the comparison. We use the default hyperparameters for the method as implemented in the Auraloss library \cite{steinmetz_auraloss_2020}. Model predictions can be found in the accompanying webpage\footnotemark[1].

The hysteresis of the models trained using the three approaches versus the target is shown in Fig.~\ref{fig:exp1b-hysteresis}. While it can be seen that the models trained using the supervised approaches match the shape of the hysteresis well, the hysteresis shape of the adversarially trained model deviates from the target. This can be explained by the adversarial model being trained on a time-frequency domain loss, which does not enforce strict time-domain matching, which is the case for the two supervised models. 

The $\mathcal{L}_\text{ESR}$ and $\mathcal{L}_\text{STFT}$ losses over the test set are listed in Table \ref{tab:exp2}, where the best (smallest) results are highlighted with bold font.
The losses for the first supervised approach are computed either via using the real delay trajectories to demodulate the targets (\textit{Demodulated}, similar to training) or applying the trajectories to the predictions (\textit{Delayed}, similar to inference).
As can be seen from the results, the second supervised approach produces the best overall performance across the considered metrics. While the time-domain loss $\mathcal{L}_\text{ESR}$ is two orders of magnitude higher for the adversarial approach than for the supervised approaches, their time-frequency domain losses $\mathcal{L}_\text{STFT}$ are of similar magnitude. For the first supervised approach, it can be seen that for both considered losses, the delayed computational method results in a smaller error. This finding suggests that demodulating the outputs can produce a larger error in comparison to delaying them with an interpolated delay line.

\subsection{Trajectory Generator}\label{ssec:experiment-1.3}

\begin{figure}[t!]
 \centering
 \adjincludegraphics[trim={0 {0.175\height} 0 0},clip]{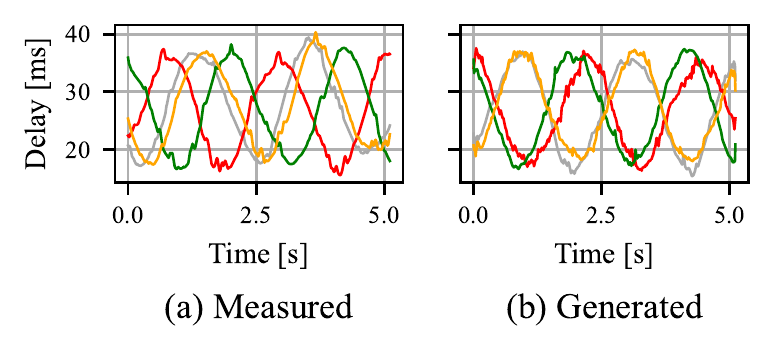}
 \vspace{-3mm}
 \caption{\label{fig:toy_trajectory_wavform}{\it Measured (left) and generated (right) delay trajectories using toy data.}}
 \vspace{-3mm}
\end{figure}

We experiment with the diffusion model approach to generate the delay trajectories from the toy experiment, which have been synthesized as described in Sec.~\ref{sec:toydata}. 
Despite our best efforts to devise a methodology for objective evaluation, we found that the stochastic nature of the data and the complexities involved made it difficult to quantify its effectiveness in a reliable and consistent manner. As such, we rely on a merely qualitative assessment, which we believe provides sufficient evidence of the successful model capabilities.
In Fig.~\ref{fig:toy_trajectory_wavform}, we show a qualitative comparison between measured trajectories and generated ones. In this case, 10 iterative steps were used to sample from the diffusion model, requiring 10 function evaluations of the neural network.
The data contains a prominent sinusoidal component with some spurious artifacts, which seem to be accurately modeled by the diffusion model. 

\section{EXPERIMENT 2: REAL DATA}\label{sec:experiment-2}

\begin{figure*}[t!]
 \centering
 \begin{subfigure}{0.5\linewidth}
  \centering
  \captionsetup{margin={7mm, 0mm}} 
  \adjincludegraphics[trim={0 {0.5\height} 0 0},clip,width=1.0\linewidth]{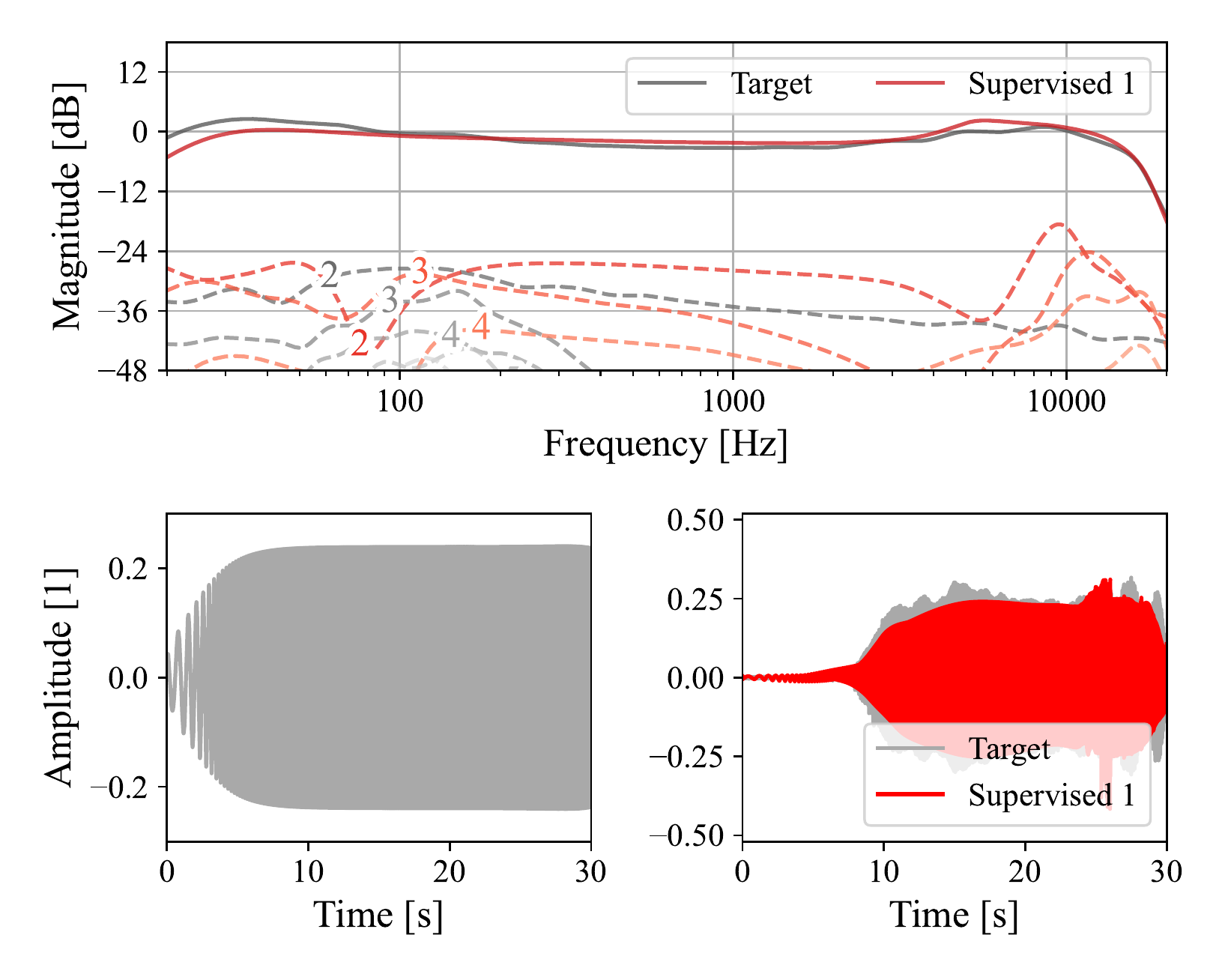}
  \vspace{-5mm}
  \caption{\label{fig:exp2-sweep-maxell-1}{\it Supervised I.}}
 \end{subfigure}%
 \begin{subfigure}{0.5\linewidth}
  \centering
  \captionsetup{margin={7mm, 0mm}} 
  \adjincludegraphics[trim={0 {0.5\height} 0 0},clip,width=1.0\linewidth]{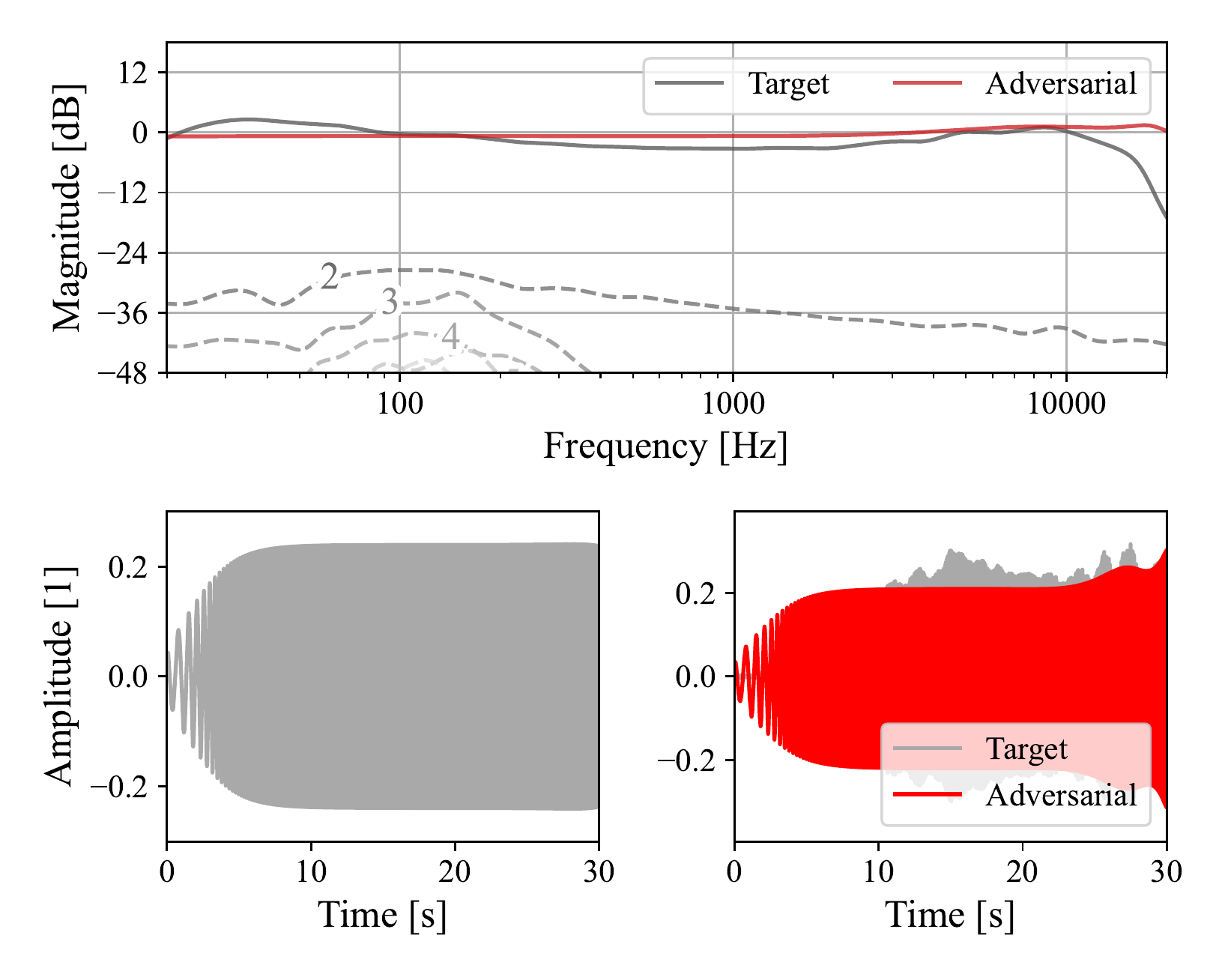}
  \vspace{-5mm}
  \caption{\label{fig:exp2-sweep-maxell-3}{\it Adversarial.}}
 \end{subfigure}
 \vspace{-5mm}
 \caption{\label{fig:exp2-sweep-maxell}{\it Model magnitude responses (solid) and distortion components (dashed), MAXELL $7 \tfrac{1}{2}$ IPS.}}
 \vspace{-3mm}
\end{figure*}

Next, the proposed method is evaluated using real data collected from the Akai 4000D tape recorder. Since the toy data did not include a noise component, this section serves as the first validation for the capability of the noise generator to learn the character of the media, as well as further validates the performance of the other two architectural components. Audio examples from the conducted experiments can be found on the web page\footnotemark[1].

\subsection{Lumped Nonlinearities and Timing Effects}
We start with the same evaluation strategy as in Sec.~\ref{ssec:experiment-1.2}, but find that the magnetic field produced by the recording head is not sufficient to saturate the considered tape formulations, as has been encountered earlier \cite{zolzer_virtual_2011}. Thus, instead of comparing the hysteresis of the trained models, we focus on the learned magnitude response and nonlinear distortion components \cite{farina_simultaneous_2000}. Only one configuration (MAXELL $7 \tfrac{1}{2}$ IPS) is evaluated here for the sake of brevity. The model predictions can be found on the accompanying web page\footnotemark[1].

We find the responses of the supervised models similar, and only show the learned magnitude response and nonlinear distortion components versus the target for the first supervised approach in Fig.~\ref{fig:exp2-sweep-maxell-1}. As can be seen, the model closely matches the shape of the linear response: the attenuated middle frequencies, the subtle emphasis of the highs, as well as the high and low-frequency roll-offs. While the target also portrays the head-bump effect in the low frequencies, this aspect is not matched by the model. We suspect that this might have to do with the dataset used for training the models not having enough low-frequency content to sufficiently learn this frequency band. While the model also learns to produce nonlinear distortion from the target data, the shape of the contours is not matched well, and the model starts to alias above about \SI{5}{\kilo\hertz}, as seen in the sudden increase of nonlinear distortion components in Fig.~\ref{fig:exp2-sweep-maxell-1}.
The learned magnitude response and nonlinear distortion components for the adversarially trained model are shown in Fig.~\ref{fig:exp2-sweep-maxell-3}. Now the match is poor in both the linear response as well as the nonlinear harmonic components. 

\begin{table}[b!]
 \vspace{-5mm}
 \caption{\itshape Real Data: Nonlinearities and timing effects.}
 \vspace{-3mm}
 \centering
 \begin{tabular}{C{0.21\columnwidth} C{0.13\columnwidth} C{0.13\columnwidth} | C{0.13\columnwidth} C{0.13\columnwidth} }
  \toprule
                & \multicolumn{2}{c|}{$\mathcal{L}_\text{ESR}$} & \multicolumn{2}{c}{$\mathcal{L}_\text{STFT}$}                              \\
  Approach      & Demod.                                        & Delayed                                       & Demod.           & Delayed \\ \midrule
  Supervised I  & $0.066$                                       & $\mathbf{0.065}$                              & $\mathbf{1.597}$ & $1.649$ \\
  Supervised II & --                                          & $0.092$                                       & --             & $1.971$ \\
  Adversarial   & --                                          & $1.193$                                       & --             & $2.437$ \\
  \bottomrule
 \end{tabular}%
 \label{tab:exp3}
\end{table}

\begin{figure}[t]
 \centering
 \adjincludegraphics[trim={0 {0.175\height} 0 {0.05\height}},clip]{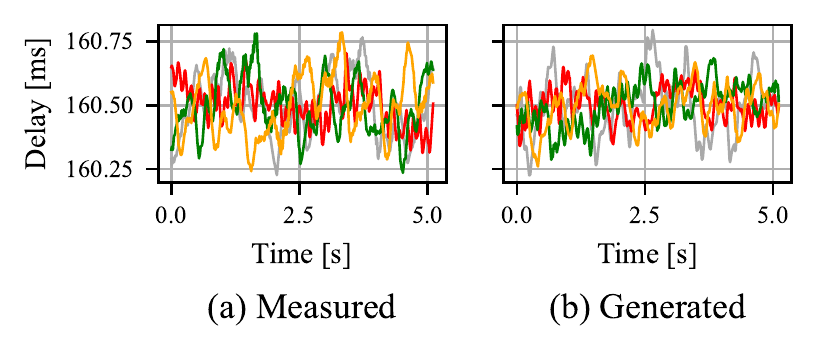}
 \vspace{-7mm}
 \caption{\it Measured (left) and generated (right) delay trajectories using real data.}
 \vspace{-2mm}
 \label{fig:delay_exp_waveforms}
\end{figure}

The $\mathcal{L}_\text{ESR}$ and $\mathcal{L}_\text{STFT}$ losses over the test set for all of the approaches are listed in Table \ref{tab:exp3}. Similarly as in Sec.~\ref{ssec:experiment-1.2}, the two supervised approaches outperform the adversarial approach in terms of both of the considered metrics, with an order of magnitude difference in the time-domain $\mathcal{L}_\text{ESR}$ loss. Unlike before, the first supervised approach performs slightly better than the second approach, although the difference is not large. While the two methods for computing the losses for the first supervised approach produce similar evaluation metrics, this time the demodulated computational method brings slight improvements in the $\mathcal{L}_\text{STFT}$ loss. This finding suggests that the error produced by the two computational methods also depends on the type of data used. 

\subsection{Trajectory Generator} \label{sec:real_delays}

\begin{figure}[t]
 \centering
 \adjincludegraphics[trim={0 {0.05\height} 0 {0.15\height}},clip,width=1.0\columnwidth]{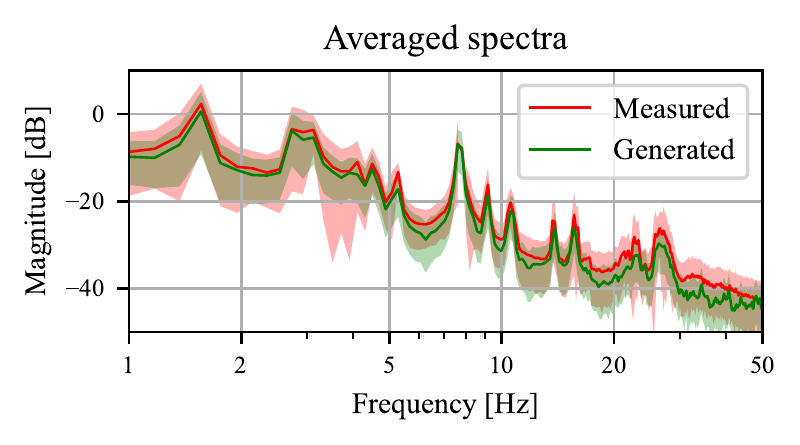}
 \vspace{-6mm}
 \caption{\it Average spectra and standard deviations of delay trajectories using real data.}
 \vspace{-2mm}
 \label{fig:delay_exp_fft}
\end{figure}

Fig.~\ref{fig:delay_exp_waveforms} shows delay trajectory samples from the measured data compared to those generated with a 10-step diffusion model.
The generated delay trajectory waveforms look realistic at first glance but, in order to provide more insights into the model behavior, we conduct a spectral analysis.
Fig.~\ref{fig:delay_exp_fft} shows the summary statistics (mean and standard deviation) of the long-term spectrum of the measured trajectories compared to that of a batch of 256 generated samples. The data presents some characteristic spectral peaks that the diffusion model is succeeding at replicating. It can also be observed that the average spectral magnitude of the generated trajectories is slightly lower than the target; we attribute this to the over-denoising phenomena that most diffusion models show, as it was observed in \cite{karras_elucidating_2022}. Nevertheless, we believe that these minor dissimilarities will pose no perceptual difference.

\subsection{Noise Generator} \label{ssec:exp_noise}

\begin{figure}
 \centering
 \adjincludegraphics[trim={0 {0.05\height} 0 {0.15\height}},clip,width=1.0\columnwidth]{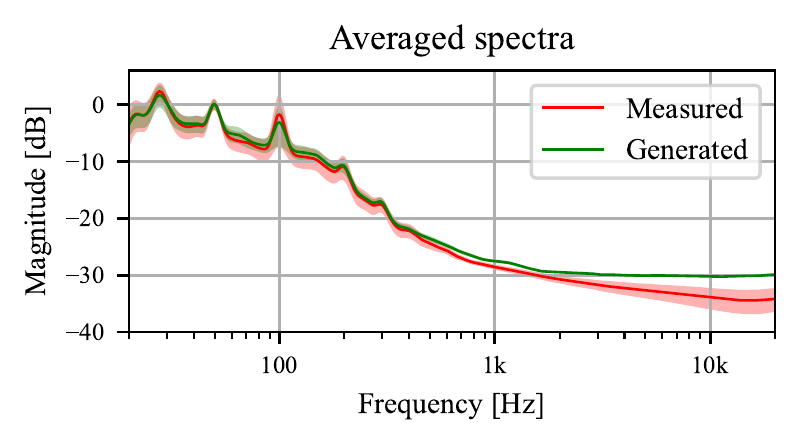}
 \vspace{-6mm}
 \caption{\it Average spectra and standard deviations of tape hiss.}
 \vspace{-2mm}
 \label{fig:noise_spectra}
\end{figure}

We assess the diffusion noise generator qualitatively, similar to the delay trajectory generator, since an objective evaluation is not feasible. Fig.~\ref{fig:noise_spectra} displays the long-term spectra of the noise data and compares it with that of the generated noises using the diffusion model with 16 discretization steps. The plot in Fig.~\ref{fig:noise_spectra} has been smoothed using a 1/6th octave band. The majority of the energy in the target data distribution is concentrated in the low-frequency region, with some spectral peaks caused by electrical noise. The diffusion model successfully replicates the spectral distribution, and the over-denoising effect observed in Sec.~\ref{sec:real_delays} does not occur as a consequence of using a stochastic sampler. 
Additionally, the noise generator can effectively model some non-stationary local characteristics in the noise data that are not adequately represented in the spectral analysis. We refer the reader to the audio examples on the companion webpage\footnotemark[1].



\section{CONCLUSIONS}\label{sec:conclusions}
This work proposed an architecture for modeling the character of magnetic tape recorders,
consisting of three components for reproducing the different aspects of the target: 1) a nonlinear block for the joint effects of the magnetic recording process as well as the record and playback amplifiers, 2) a time-varying delay line controlled by a delay trajectory generator for imperfections in the tape transport, and 3) a noise generator for the tape hiss. The different blocks were implemented using separate neural network architectures: an RNN for the nonlinear block and separate diffusion models for the delay and noise generators. Three training schemes were considered for the nonlinear block: two supervised and one adversarial.

Our results indicate that the RNN architecture is suitable for learning the characteristic hysteretic nonlinear behavior of the tape magnetization. This was also found recently elsewhere for the related case of audio transformers \cite{massi_deep_2023}.
Based on objective and qualitative evaluation and informal listening, the two supervised approaches for the nonlinear block together with the generative models for the delay trajectories and tape hiss capture the perceptual character of the tape recorder as a whole. While the proposed supervised training schemes require the target machine to operate in stereo, an aspect of which was circumvented by the adversarial approach, this latter approach did not prove mature yet in learning the nonlinear character of the tape.


\section{Acknowledgments}
We acknowledge Aalto Science IT for the computational resources.

\bibliographystyle{IEEEbib}
\bibliography{2023_DAFX_Neural_Tape}

\end{document}